\documentclass[aps,prl,twocolumn,groupedaddress]{revtex4-1}

\usepackage{graphicx}
\usepackage{bm}
\usepackage{amsmath}
\usepackage{subfigure}
\usepackage{hyperref}

\begin{document}

\title{Dynamic Theory of Polydomain Liquid-Crystal Elastomers }
\author{Ayhan Duzgun}
\author{Jonathan V. Selinger}
\affiliation{Liquid Crystal Institute, Kent State University, Kent, OH 44242}

\date{February 25, 2015}

\begin{abstract}
When liquid-crystal elastomers are prepared without any alignment, disordered polydomain structures emerge as the materials are cooled into the nematic phase.  These polydomain structures have been attributed to quenched disorder in the cross-linked polymer network.  As an alternative explanation, we develop a theory for the dynamics of the isotropic-nematic transition in liquid-crystal elastomers, and show that the dynamics can induce a polydomain structure with a characteristic length scale, through a mechanism analogous to the Cahn-Hilliard equation for phase separation.
\end{abstract}

\maketitle

Liquid-crystal elastomers are remarkable materials that combine the elastic properties of cross-linked polymer networks with the anisotropy of liquid crystals~\cite{Warner2003}.  Any distortion of the polymer network affects the orientational order of the liquid crystal, and any change in the magnitude or direction of liquid-crystal order influences the shape of the polymer network.  Hence, these elastomers are useful for applications as actuators or shape-changing materials.

For many applications, it is necessary to prepare monodomain liquid-crystal elastomers.  In practice, this can be done by applying a mechanical load or other aligning field while crosslinking~\cite{Kupfer1991}.  Surprisingly, elastomers prepared without an aligning field do not form monodomains.  Rather, they form polydomain structures with nematic order in local regions, which are macroscopically disordered.  These polydomain structures have been seen in many experiments, using a wide range of techniques~\cite{Clarke1998,Elias1999,Urayama2007,Feio2008,Urayama2009}.  Indeed, a recent polarized light scattering study shows that liquid-crystal elastomers evolve toward a state of increasing disorder as the isotropic-nematic transition proceeds, unless the disorder is suppressed by a gradually increasing load~\cite{Grabowski2011}.

One important issue in the theory of liquid-crystal elastomers is how to understand the polydomain state.  Several theoretical studies have attributed this state to quenched disorder in the polymer network, which can be understood by analogy with spin glass theory~\cite{Fridrikh1997,Fridrikh1999,Yu1999,Xing2003,Petridis2006,Xing2008,Lu2012,Xing2013,Lu2013}.  Effects of quenched disorder have further been modeled and visualized through numerical simulations~\cite{Yu1998,Uchida1999,Uchida2000,Selinger2004}.  More macroscopic theories have shown that the resulting polydomain structure has profound consequences for the material's effective elasticity~\cite{Biggins2009,Biggins2012}.

\begin{figure}
(a)\includegraphics[width=.95\columnwidth]{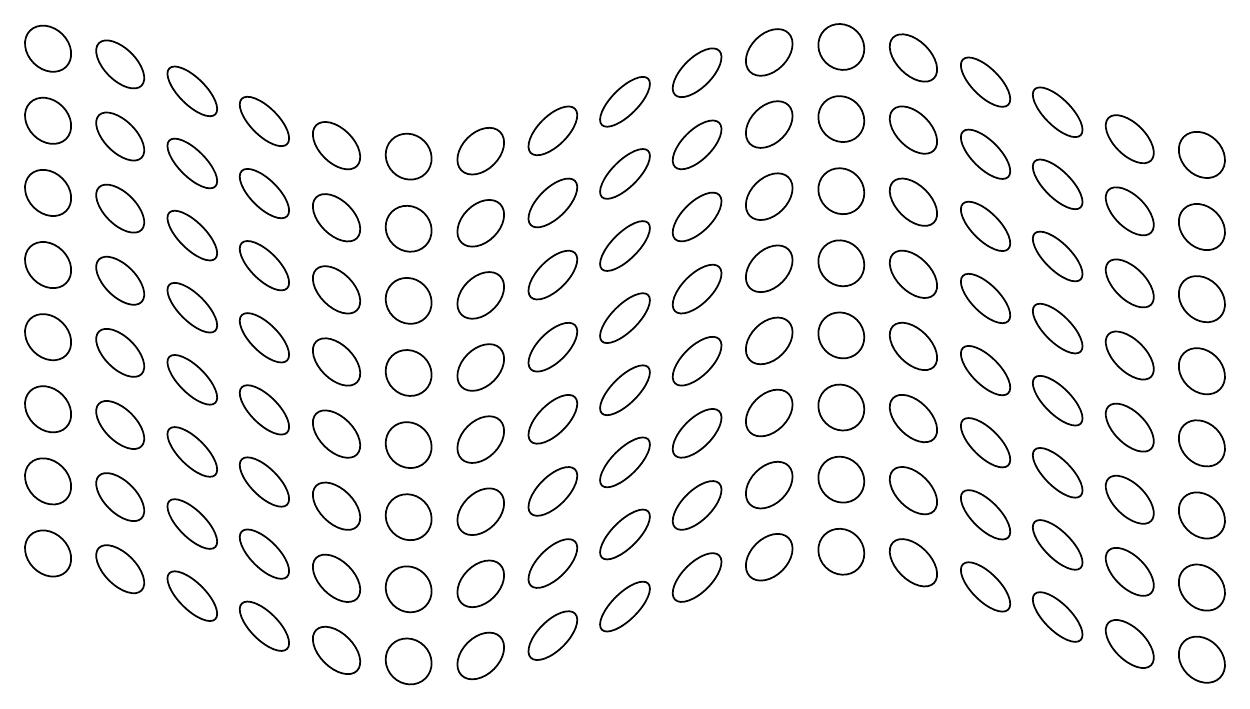}
(b)\includegraphics[width=.95\columnwidth]{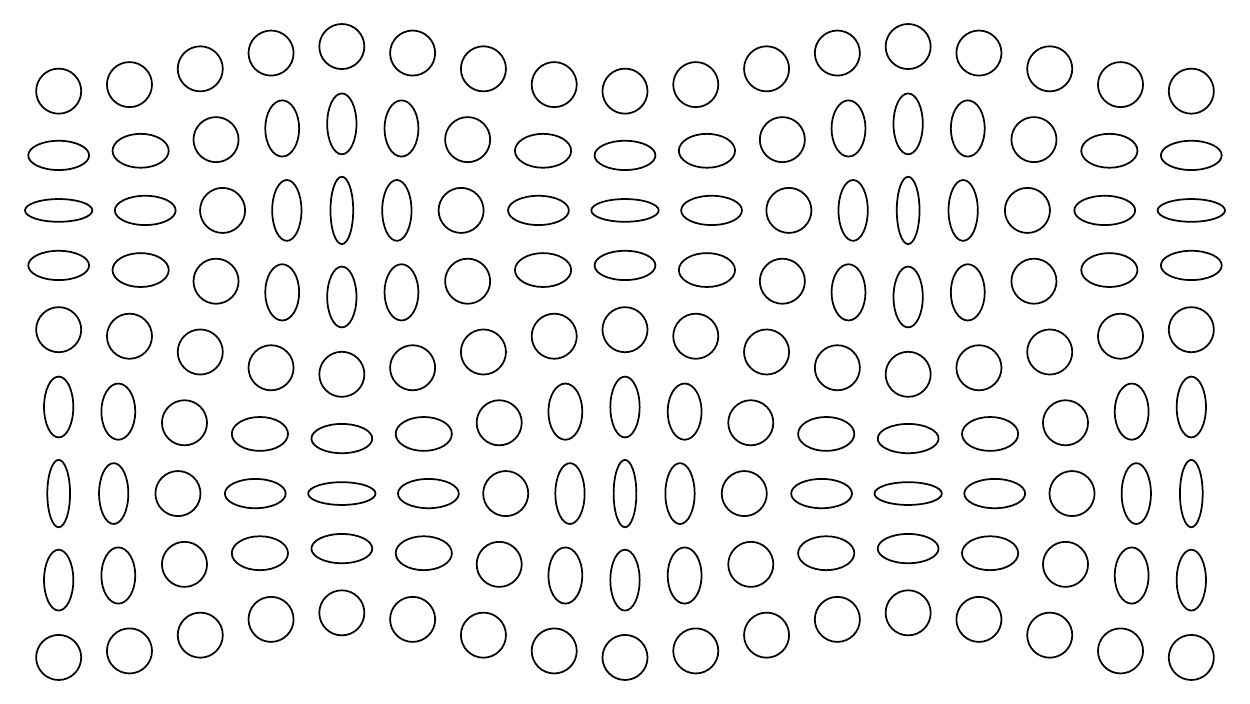}
(c)\includegraphics[width=.95\columnwidth]{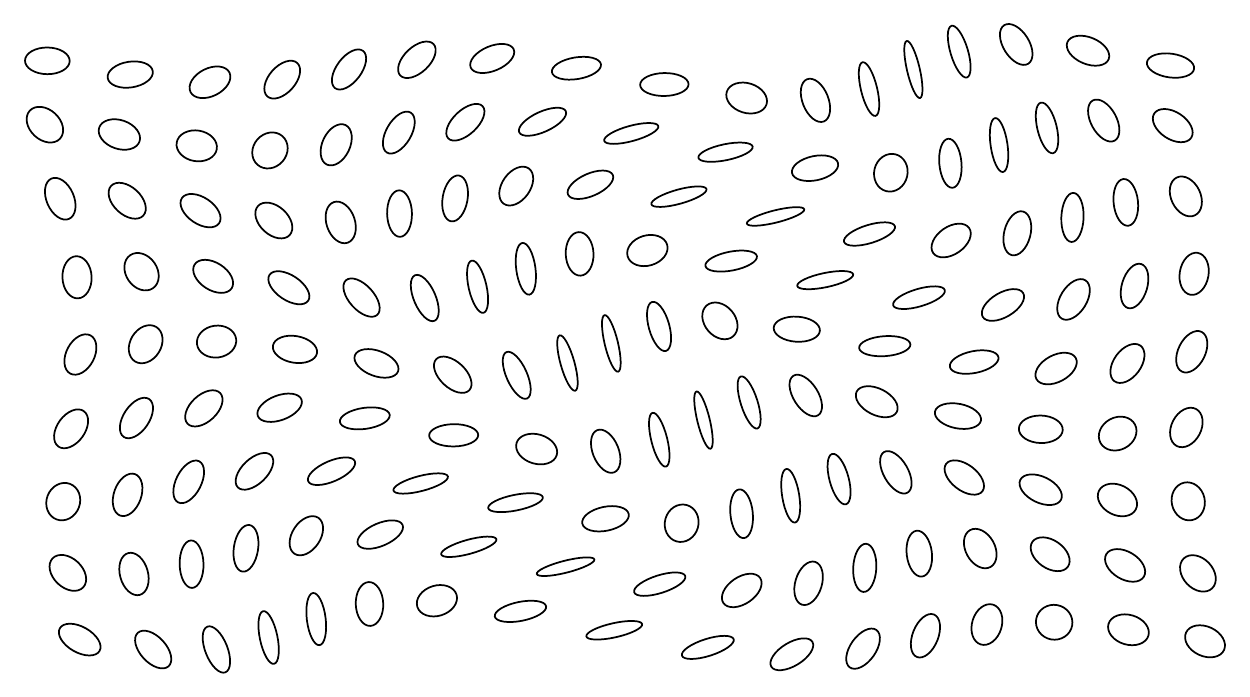}
\caption{Visualization of liquid-crystal elastomer structures calculated here.  The orientation and eccentricity of ellipses represents local nematic order.  (a)~Single wave in nematic order and displacement.  (b)~Superposition of two perpendicular waves, forming a square lattice.  (c)~Superposition of three waves with random directions, amplitudes, and phases.}
\label{visualization}
\end{figure}

The purpose of this paper is to suggest a different mechanism for the origin of the polydomain state, not related to quenched disorder.  We develop a theory for the dynamics of the isotropic-nematic transition in liquid-crystal elastomers, in which growing nematic order is coupled to elastic strain.
We explore this theory in two dimensions (2D), using two models for dynamic evolution of nematic order and strain.  The theory shows that dynamics can itself select a characteristic length scale for a disordered polydomain structure, through a mechanism similar to the Cahn-Hilliard equation for phase separation.  In particular, the theory predicts formation of structures with the form shown in Fig.~\ref{visualization}.  We suggest that this mechanism may play a role in formation of polydomain liquid-crystal elastomers, in addition to quenched disorder.

In the theory of 2D liquid-crystal elastomers, nematic order is described by the tensor order parameter $Q_{\alpha\beta}(\bm{r})$, and elastic distortion of the material by the displacement vector $\bm{u}(\bm{r})$.  In terms of displacement, the strain tensor is defined as $\epsilon_{\alpha\beta}=\frac{1}{2}[\partial_\alpha u_\beta + \partial_\beta u_\alpha + (\partial_\alpha u_\gamma)(\partial_\beta u_\gamma)]$.  The free energy can be expressed in terms of $Q_{\alpha\beta}$ and $\epsilon_{\alpha\beta}$ as
\begin{align}
F=\int d^2 r \biggl[&
\frac{1}{2}a Q_{\alpha\beta}Q_{\alpha\beta}
+\frac{1}{4}b (Q_{\alpha\beta}Q_{\alpha\beta})^2\nonumber\\
&+\frac{1}{2}L (\partial_\gamma Q_{\alpha\beta})(\partial_\gamma Q_{\alpha\beta})
+\frac{1}{2}\lambda \epsilon_{\alpha\alpha}\epsilon_{\beta\beta}\nonumber\\
&+\mu \epsilon_{\alpha\beta}\epsilon_{\alpha\beta}
-V \epsilon_{\alpha\beta} Q_{\alpha\beta}
\biggr].
\label{freeenergy}
\end{align}
Here, the first two terms are the Landau-de Gennes expansion for the free energy in powers of the order tensor.  The coefficient $a=a'(T-T_0)$ is assumed to vary linearly with temperature, while $b$ is a positive constant.  The third term is the Frank free energy for spatial variations in the order tensor, assuming a single Frank coefficient $L$.  The fourth and fifth terms are the elastic free energy in terms of the strain tensor, with Lam\'e coefficients $\lambda$ and $\mu$.  The final term is the coupling between nematic order and strain, with coefficient $V$.

If there were no coupling between nematic order and strain, $V=0$, the system would have an isotropic-nematic transition at $a=0$, corresponding to temperature $T_0$.  With coupling $V\not=0$, the transition is shifted upward to $a=V^2 /(2\mu)$, corresponding to the higher temperature $T_\text{IN}=T_0 + V^2 /(2\mu a')$.  Above that temperature, the state of minimum free energy is uniformly isotropic, with $Q_{\alpha\beta}=0$ and $\epsilon_{\alpha\beta}=0$.  Below that temperature, at $a=V^2 /(2\mu)-\delta a$, the state of minimum free energy becomes uniformly nematic, with alignment along a randomly selected director $\hat{\bm{n}}$.  In this state, the order tensor is $Q_{\alpha\beta}=S(2 n_\alpha n_\beta -\delta_{\alpha\beta})$, where the magnitude of nematic order is $S=\sqrt{\delta a/(2b)}$.  This state extends uniformly along the director, with strain $\epsilon_{\alpha\beta}=[V/(2\mu)]Q_{\alpha\beta}$.

Now suppose we begin in the isotropic phase, and rapidly cool to a temperature slightly below $T_\text{IN}$.  At this low temperature, nematic order and strain both begin to grow dynamically.  We ask:  Does the dynamic process lead to the state of minimum free energy, with uniform $Q_{\alpha\beta}$ and $\epsilon_{\alpha\beta}$?  Alternatively, does it lead to a different, nonuniform state?

To answer this question, we develop a model for the dynamics of the phase transition.  We actually consider two models, first simple linear drag and then more realistic viscous flow.  In both models, we describe four coupled degrees of freedom:  $Q_{xx}(\bm{r},t)$, $Q_{xy}(\bm{r},t)$, $u_x(\bm{r},t)$, and $u_y(\bm{r},t)$.  The remaining components of $Q_{\alpha\beta}(\bm{r},t)$ are fixed because it is a symmetric, traceless tensor, and  $\epsilon_{\alpha\beta}(\bm{r},t)$ can be derived from $\bm{u}(\bm{r},t)$.  We cannot take the strain tensor components as our fundamental degrees of freedom because they are constrained by elastic compatibility; they must all be derivable from the same $\bm{u}(\bm{r},t)$.

In the simplest model of overdamped dynamics with linear drag, the rate of change for each degree of freedom is linearly proportional to the force acting on it.  Hence, the equations of motion are
\begin{align}
&\frac{\partial Q_{xx}}{\partial t}= -\Gamma_Q \frac{\delta F}{\delta Q_{xx}},\quad
\frac{\partial Q_{xy}}{\partial t}= -\Gamma_Q \frac{\delta F}{\delta Q_{xy}},\nonumber\\
&\frac{\partial u_x}{\partial t}= -\Gamma_u \frac{\delta F}{\delta u_x},\quad
\frac{\partial u_y}{\partial t}= -\Gamma_u \frac{\delta F}{\delta u_y},
\end{align}
where $\Gamma_Q$ and $\Gamma_u$ are mobility coefficients.  To calculate the forces on the right side of those equations, we substitute the definition of the strain tensor into the free energy~(\ref{freeenergy}), and take functional derivatives with respect to $Q_{\alpha\beta}$ and $u_\alpha$.  We then linearize the equations, assuming that $Q_{\alpha\beta}$ and $u_\alpha$ are both small in early stages of nematic ordering.  The equations then become
\begin{align}
&\frac{\partial Q_{xx}}{\partial t}= \Gamma _Q \left[-2 a Q_{xx}+2 L \nabla^2 Q_{xx}
+V(\partial_x u_x-\partial_y u_y)\right],\nonumber\\
&\frac{\partial Q_{xy}}{\partial t}= \Gamma _Q \left[-2 a Q_{xy}+2 L \nabla^2 Q_{xy}
+V(\partial_x u_y+\partial_y u_x)\right],\nonumber\\
&\frac{\partial u_x}{\partial t}= \Gamma _u \left[(\lambda+\mu)\partial_x \bm{\nabla}\cdot\bm{u}
+\mu \nabla^2 u_x
-V (\bm{\nabla}\cdot\bm{Q})_x \right],\nonumber\\
&\frac{\partial u_y}{\partial t}= \Gamma _u \left[(\lambda+\mu)\partial_y \bm{\nabla}\cdot\bm{u}
+\mu \nabla^2 u_y
-V (\bm{\nabla}\cdot\bm{Q})_y \right].
\end{align}

To simplify this system of equations, we Fourier transform from position $\bm{r}$ to wavevector $\bm{k}$, then write the equations in the matrix form
\begin{equation}
\frac{\partial}{\partial t}
\begin{pmatrix}
Q_{xx}(\bm{k},t) \\
Q_{xy}(\bm{k},t) \\
u_x(\bm{k},t) \\
u_y(\bm{k},t) \\
\end{pmatrix}
=-M(\bm{k})
\begin{pmatrix}
Q_{xx}(\bm{k},t) \\
Q_{xy}(\bm{k},t) \\
u_x(\bm{k},t) \\
u_y(\bm{k},t) \\
\end{pmatrix},
\label{matrixeq}
\end{equation}
where $M(\bm{k})$ is a $4\times4$ matrix.  This matrix equation resembles the Cahn-Hilliard equation for phase separation of a binary fluid.  At each $\bm{k}$, the matrix $M(\bm{k})$ has four eigenmodes $i$, which either grow or decay exponentially as $e^{-\Lambda_i(\bm{k})t}$, where $\Lambda_i(\bm{k})$ is the corresponding eigenvalue of $M(\bm{k})$.  Note that $\Lambda_i(\bm{k})<0$ corresponds to exponential growth, while $\Lambda_i(\bm{k})>0$ corresponds to exponential decay.  We must determine what grows most rapidly:  which eigenmode at which wavevector?

To identify the fastest-growing mode, we choose coordinates such that $\bm{k}$ is along the $x$-axis.  The matrix then simplifies to
\begin{align}
&M(\bm{k})=\\
&\setlength{\arraycolsep}{0pt}\begin{pmatrix}
2\Gamma_Q(a+Lk^2) & 0 & -i\Gamma_Q V k & 0 \\
0 & 2\Gamma_Q(a+Lk^2) & 0 & -i\Gamma_Q V k \\
i\Gamma_u V k & 0 & \Gamma_u(\lambda+2\mu)k^2 & 0 \\
0 & i\Gamma_u V k & 0 & \Gamma_u \mu k^2 \\
\end{pmatrix}.\nonumber
\end{align}
We now take the limit of an incompressible material, with $\lambda\to\infty$.  In this limit, $u_x$ has a high energy cost, so that it decays rapidly, and hence we eliminate it from consideration.  In that case, $Q_{xx}$ is not coupled to any other degrees of freedom, so it is an eigenmode by itself, with eigenvalue $2\Gamma_Q(a+Lk_x^2)$.  If the system is at a temperature slightly below the isotropic-nematic transition, we must have $0<a<V^2 /(2\mu)$.  In that temperature range, this eigenvalue is positive, so that $Q_{xx}$ decays exponentially.  Hence, we also eliminate it from consideration in the search for the fastest-growing mode.

\begin{figure}
\includegraphics[width=\columnwidth]{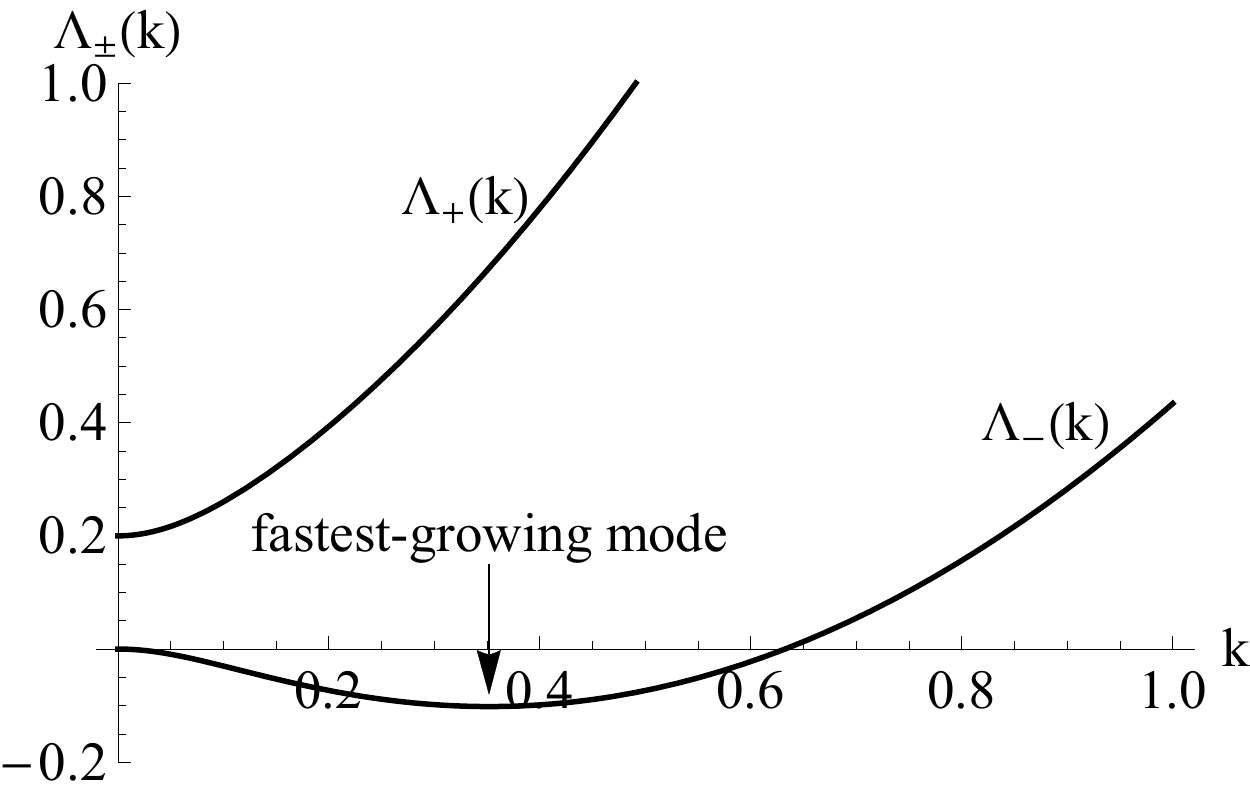}
\caption{Sample plot of the eigenvalues $\Lambda_\pm(\bm{k})$ in the linear drag model of dynamics, with parameters $a=0.1$ and $L=\mu=V=\Gamma_Q=\Gamma_u=1$.  The largest negative eigenvalue corresponds to the fastest-growing mode, which occurs at a dynamically selected wavevector.}
\label{drag2modes}
\end{figure}

The remaining two modes are linear combinations of $Q_{xy}$ and $u_y$, with eigenvalues
\begin{align}
\Lambda_\pm(k)=& \Gamma_Q(a+L k^2)+\textstyle{\frac{1}{2}}\Gamma_u \mu k^2 \\
&\pm\sqrt{\left[\Gamma_Q(a+L k^2)-\textstyle{\frac{1}{2}}\Gamma_u \mu k^2\right]^2 +\Gamma_Q \Gamma_u V^2 k^2} \nonumber
\end{align}
Figure~\ref{drag2modes} shows a sample plot of these two eigenvalues as functions of $k$.  The eigenvalue $\Lambda_+(k)$ begins at $2\Gamma_Q a$ when $k=0$, then increases with increasing $k$.  For temperatures just below the isotropic-nematic transition, with $0<a<V^2 /(2\mu)$, it is always positive and hence represents a decaying mode.  By contrast, $\Lambda_-(k)$ begins at $0$ when $k=0$, then decreases into negative values over the range $0<k<\sqrt{\delta a/L}$, where $\delta a=V^2 /(2\mu)-a$, and eventually returns to positive values for larger $k$.  Over the range in which it is negative, it represents an exponentially growing mode.  To find the fastest-growing wavevector, we minimize $\Lambda_-(k)$ over $k$.  For temperatures just below the isotropic-nematic transition, for small $\delta a$, this wavevector is $k_\text{fastest}\approx\sqrt{\delta a/(2L)}$, and the corresponding growth rate is $|\Lambda_-(k_\text{fastest})|\approx\Gamma_u \mu^2 \delta a^2/(2 L V^2)$.

We emphasize that this wavevector is selected through a dynamic mechanism.  It is not the minimum of the free energy (which is a state of uniform nematic order and strain).  Moreover, it only occurs because of the coupling $V$ between nematic order and strain in a liquid-crystal elastomer.  If these variables were uncoupled ($V=0$), the matrix $M$ would be diagonal, the isotropic-nematic transition would occur at $a=0$, and the fastest-growing mode below that transition would be $k=0$.

To characterize the fastest-growing mode, we calculate the eigenvector of $M$ corresponding to eigenvalue $\Lambda_-$ at wavevector $k_\text{fastest}$.  This eigenvector represents waves in both $Q_{xy}$ and $u_y$ (with our assumption that the wavevector is in the $x$-direction), and these waves are $90^\circ$ out of phase.  Figure~\ref{visualization}(a) shows a sample visualization of the structure with a single Fourier mode.  It has alternating stripes with the director oriented at $\pm45^\circ$ from the wavevector, accompanied by displacement perpendicular to the wavevector.

In general, a liquid-crystal elastomer will not have only one Fourier mode.  Rather, it can include modes with wavevectors of magnitude $k_\text{fastest}$ in multiple directions.  To find a mode in an arbitrary direction, we rotate the wavevector, and make a corresponding rotation of $Q_{\alpha\beta}$ and $\bm{u}$.  We then add up the Fourier modes to find the structure.  Figure~\ref{visualization}(b) shows an example with two perpendicular waves of equal amplitude, leading to a square lattice in the nematic order and the displacement.  Figure~\ref{visualization}(c) shows a more realistic example with a superposition of three waves with random directions, amplitudes, and phases.

The structures in Fig.~\ref{visualization} are similar to structures commonly observed in experiments and simulations on \emph{active} nematic liquid crystals~\cite{Marchetti2013}.  This similarity is reasonable, because both systems are controlled by couplings between orientational order and extension of the material.

The growth of nematic order in a liquid-crystal elastomer can be described by the dynamic correlation function
\begin{align}
&C(|\bm{r}-\bm{r}'|,t)=\left\langle\cos2[\theta(\bm{r})-\theta(\bm{r}')]\right\rangle_t\nonumber\\
&=\left\langle Q_{xx}(\bm{r},t)Q_{xx}(\bm{r}',t)+Q_{xy}(\bm{r},t)Q_{xy}(\bm{r}',t)\right\rangle\nonumber\\
&=\sum_{\bm{k}} e^{i\bm{k}\cdot(\bm{r}-\bm{r}')} \left\langle |Q_{xx}(\bm{k},t)|^2 + |Q_{xy}(\bm{k},t)|^2\right\rangle.
\end{align}
This sum is dominated by the fastest-growing mode at wavevectors with magnitude $k_\text{fastest}$, and hence
\begin{align}
C(|\bm{r}-\bm{r}'|,t)
&\propto\sum_{|\bm{k}|=k_\text{fastest}} e^{i\bm{k}\cdot(\bm{r}-\bm{r}')} e^{2|\Lambda_-(k_\text{fastest})|t}\nonumber\\
&\propto J_0(k_\text{fastest}|\bm{r}-\bm{r}'|)e^{2|\Lambda_-(k_\text{fastest})|t}.
\end{align}
Thus, in the early stages of growth, the correlation function has the form of Bessel function $J_0(k_\text{fastest}|\bm{r}-\bm{r}'|)$, with an exponentially increasing magnitude.  In later stages of growth, the approximation of small nematic order ceases to apply, and other types of modeling are needed.  Even so, the length scale of $1/k_\text{fastest}$ is established from the early stages.

The dynamic model presented above has a limitation:  It assumes that both $Q_{\alpha\beta}(\bm{r},t)$ and $\bm{u}(\bm{r},t)$ have overdamped dynamics, with drag forces linearly proportional to the rate of change of these quantities.  This assumption is appropriate for dynamics on a substrate, where the dissipation is caused by drag against the substrate.  However, if there is no substrate, it is reasonable to generalize the dynamics in two ways:  by considering inertia for the displacement and by considering viscous dissipation rather than drag against a substrate.

For this generalization, we use the equations of motion
\begin{align} 
\rho\frac{\partial^2 u_\alpha}{\partial t^2}=&-\frac{\delta D}{\delta\dot{u}_\alpha}-\frac{\delta F}{\delta u_\alpha},\nonumber\\
0=&-\frac{\delta D}{\delta\dot{Q}_{\alpha\beta}}-\frac{\delta F}{\delta Q_{\alpha\beta}}.
\end{align} 
Here $\rho$ is the mass density, which gives inertia for $\bm{u}$; there is no inertia for $Q_{\alpha\beta}$.  Also, $D$ is the Rayleigh dissipation function, which can be written as
\begin{equation}
D=\int d^2 r \left[\eta A_{\alpha\beta} A_{\alpha\beta}+\frac{1}{2}\gamma_1 B_{\alpha\beta} B_{\alpha\beta}
+\gamma_2 A_{\alpha\beta} B_{\alpha\beta}\right]
\end{equation}
in terms of the two modes that dissipate energy:  $A_{\alpha\beta}=\frac{1}{2}(\partial_\alpha\dot u_\beta+\partial_\beta\dot u_\alpha)$ is the rate of shear flow, and $B_{\alpha\beta}=\dot Q_{\alpha\beta}-\omega_z(\epsilon_{\delta\alpha}Q_{\delta\beta}+\epsilon_{\delta\beta}Q_{\delta\alpha})$ is the rotation rate of nematic order relative to rotational flow of the material, given by $\omega_z=\frac{1}{2}\epsilon_{\mu\nu}\partial_\mu\dot u_\nu$.  In these expressions, $\eta$ is the viscosity, $\gamma_1$ is the rotational viscosity, and $\gamma_2$ is a dissipative coupling coefficient.

We combine these expressions to derive the coupled equations of motion for $Q_{xx}$, $Q_{xy}$, $u_x$, and $u_y$, and linearize the equations assuming these variables are small in the early stages of nematic ordering.  We then follow the same steps as in the previous calculation:  Fourier transform from $\bm{r}$ to $\bm{k}$, choose coordinates such that $\bm{k}$ is along the $x$-axis, eliminate $u_x$ by the constraint of incompressibility, and eliminate $Q_{xx}$ because it is an independent, exponentially decaying mode.  We are left with a matrix equation for $Q_{xy}(\bm{k},t)$ and $u_y(\bm{k},t)$,
\begin{align}
\begin{pmatrix}
0 & 0\\
0 & \rho
\end{pmatrix}
\begin{pmatrix}
\ddot{Q}_{xy} \\
\ddot{u}_{y}
\end{pmatrix}=
&-\begin{pmatrix}
4\gamma_1 & i\gamma_2 k\\
-i\gamma_2 k & \eta k^2
\end{pmatrix}
\begin{pmatrix}
\dot{{Q}}_{xy} \\
\dot{{u}}_{y}
\end{pmatrix}\\
&-\begin{pmatrix}
2(a+L k^2) & -i V k \\
i V k & \mu k^2
\end{pmatrix}
\begin{pmatrix}
Q_{xy} \\
u_{y}
\end{pmatrix}.\nonumber
\end{align}
Next we Fourier transform from time $t$ to frequency $\omega$, and obtain
\begin{equation}
\begin{pmatrix}
2(a+L k^2)-4 i\gamma_1\omega & -iVk+\gamma_2\omega k\\
iVk-\gamma_2\omega k & \mu k^2-i\eta\omega k^2-\rho\omega^2\\
\end{pmatrix}
\begin{pmatrix}
Q_{xy} \\
u_{y}
\end{pmatrix}
=0.
\end{equation}

In this matrix equation, there are two couplings between $Q_{xy}$ and $u_y$:  the elastic coupling $V$ and the dissipative coupling $\gamma_2$.  For simplicity, we set $\gamma_2=0$ and consider only the elastic coupling.

The matrix equation only allows nontrivial $Q_{xy}$ and $u_y$ if the determinant of the matrix is zero.  Hence, we set the determinant to zero and solve for the allowed frequencies $\omega$.  Because the determinant is a cubic function of $\omega$, there are three solutions.  Expanding to first order in $1/\rho$, the solutions are
\begin{align}
&\omega_0 (k) = -\frac{i (a+Lk^2)}{2\gamma_1} +\frac{i\gamma_1 V^2 k^2}{\rho (a+Lk^2)^2},\\
&\omega_\pm (k) = \pm k\sqrt{\frac{\mu}{\rho}-\frac{V^2}{2\rho(a+Lk^2)}} -\frac{i k^2}{2\rho}\left[\eta+\frac{\gamma_1 V^2}{(a+Lk^2)^2}\right] .\nonumber
\end{align}
Here, a real part of $\omega$ represents oscillation, a negative imaginary part represents exponential decay, and a positive imaginary part represents exponential growth.  

The solution $\omega_0 (k)$ is a purely damped mode.  Whether the system is in the isotropic phase, $a>V^2/(2\mu)$, or slightly in the nematic phase, $0<a<V^2/(2\mu)$, this mode decays exponentially.

The modes $\omega_\pm (k)$ depend on whether system is in the isotropic or nematic phase.  In the isotropic phase, $a>V^2/(2\mu)$, these modes are damped sound waves, with both oscillation and exponential decay.  By comparison, when the system is cooled slightly into the nematic phase, $0<a<V^2/(2\mu)$, these modes change into pure exponential growth or decay.  One of the modes has a negative imaginary part for all $k$, corresponding to decay, but the other mode has a positive imaginary part for a range of $k$, corresponding to growth.

\begin{figure}
\includegraphics[width=\columnwidth]{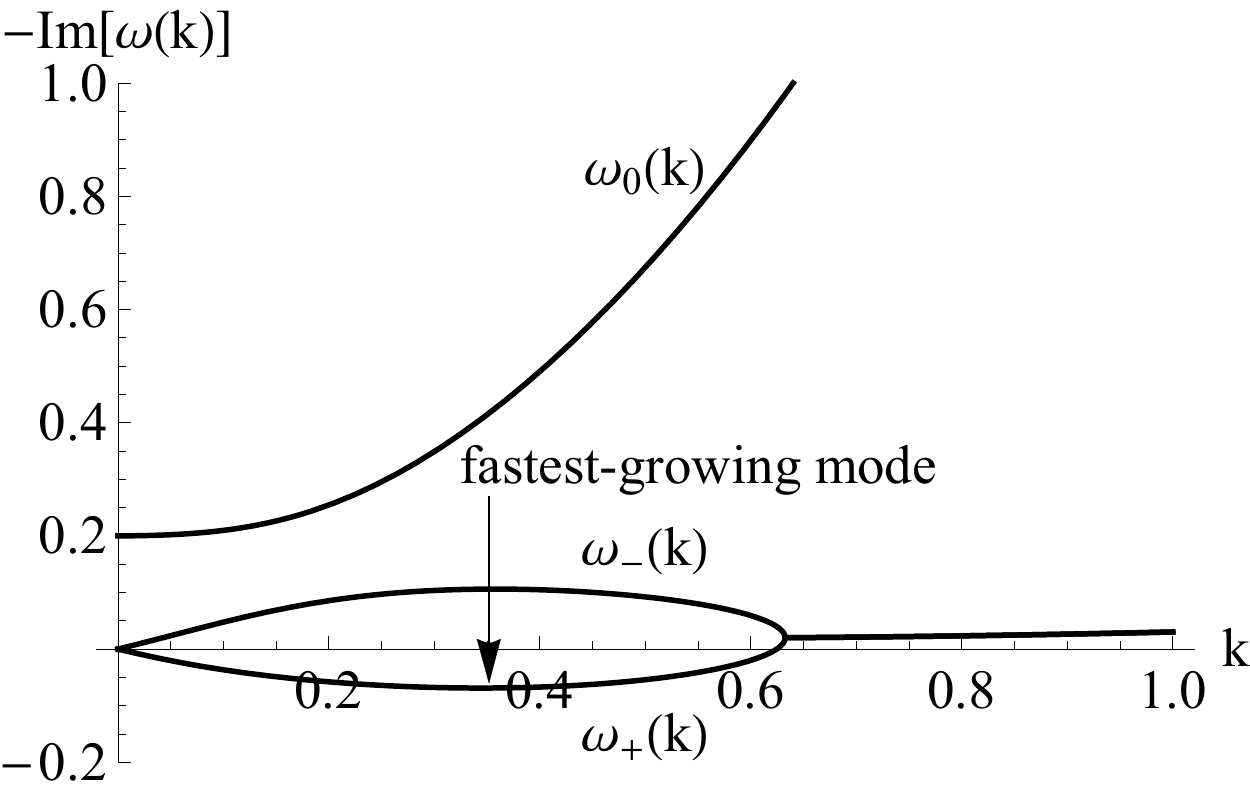}
\caption{Sample plot of the mode structure in the generalized model of dynamics, with inertia and viscosity.  Parameters are $a=0.1$, $L=\mu=V=\eta=1$, $\gamma_1=0.25$, and $\rho=20$.  The quantity $-\mathrm{Im}[\omega(k)]$ is the exponential decay rate, equivalent to $\Lambda_{\pm}(k)$ in Fig.~\ref{drag2modes}.  The largest \emph{positive} value of $\mathrm{Im}[\omega(k)]$ corresponds to the fastest-growing mode.}
\label{viscous3modes}
\end{figure}

Figure~\ref{viscous3modes} shows a sample plot of the mode structure in the nematic phase.  We can see that the $\omega_0 (k)$ and $\omega_- (k)$ modes are decaying for all $k$, but the $\omega_+ (k)$ mode is growing for a range of $k$.  In this respect, it resembles the growing mode in the linear drag model of dynamics, shown in Fig.~\ref{drag2modes}.  In the limit of high $\rho$, the range of exponential growth is $0<k<\sqrt{\delta a/L}$, and the fastest-growing wavevector is $k_\text{fastest}\approx\sqrt{\delta a/(2L)}$, where $\delta a=V^2 /(2\mu)-a$.  These results are equivalent to corresponding results for the linear drag model.

Hence, the generalized model of dynamics (with inertia and viscosity) leads to the same conclusion as the linear drag model:  The dynamic mechanism of the isotropic-nematic transition selects a fastest-growing wavevector.  This fastest-growing wavevector is not the minimum of the free energy, and it only occurs because of the coupling between nematic order and strain.  We expect modulations with this wavevector to grow in liquid-crystal elastomers cooled below the isotropic-nematic transition, leading to structures with the form shown in Fig.~\ref{visualization}.

To be sure, both models of dynamics presented here apply only to early stages of growth of nematic order.  In later stages, as nematic order becomes more established, we cannot assume that $Q_{\alpha\beta}(\bm{r},t)$ and $\bm{u}(\bm{r},t)$ are small.  In that case, our linearization of the equations of motion breaks down, and the dynamics must be studied through other techniques, such as numerical simulation.  Hence, we cannot be sure whether the polydomain structure will persist into longer time, or will eventually coarsen into a uniform structure.  In those stages of dynamics, quenched disorder may lock in the polydomain structure at the length scale given by dynamics, and prevent it from coarsening away.

In conclusion, we have shown that dynamic evolution of nematic order can induce a polydomain state with a characteristic length scale, in the early stages of the isotropic-nematic transition.  This mechanism should be considered, along with quenched disorder, in studies of polydomain liquid-crystal elastomers.

We thank D. R. Nelson and M. Y. Pevnyi for helpful discussions.  This work was supported by NSF Grant DMR-1409658.

\bibliography{LCEtheory2}

\end{document}